# Maximizing the sensitivity of ELT to habitable worlds with a space-based starshade

*A white paper in response to the ESO Expanding Horizons initiative*

**Corresponding author:**
Markus Janson, Stockholm University, Stockholm, Sweden, markus.janson@astro.su.se

**Co-authors:**
Enric Palle, Instituto Astrofisica de Canarias, La Laguna, Spain
Thomas Henning, Max Planck Institute for Astronomy, Heidelberg, Germany
Sascha P. Quanz, Swiss Federal Institute of Technology, Zurich, Switzerland
Lars Buchhave, Technical University of Denmark, Kongens Lyngby, Denmark
Alexis Brandeker, Stockholm University, Stockholm, Sweden

**Abstract:**

The ELT will provide groundbreaking science across a wide range of areas, including small habitable-zone exoplanets; however, true Earth analogs in the habitable zones of Sun-like stars are generally beyond the reach even of the ELT, due to the extreme contrast ratio and small angular separation between the planet and star. Here, we note that the combination of ELT and a space-based starshade would provide the contrast needed to observe potentially tens of Earth analogs, as well as other planets. This would yield the scientific basis needed for addressing central scientific questions regarding the frequency and distribution of habitability and life in the Universe. The huge aperture of ELT, combined with a contrast otherwise only reachable in space, opens up scientific avenues that are unmatched by any other existing or foreseen facility. ESO could conceivably collaborate with ESA (and others) to facilitate a starshade mission suitable for synergy with the ELT, as well as to prepare the ELT instrumentation in order to maximize its potential for synergy with a starshade.

**Habitable planets and the ELT**

The prospect to detect and characterize Earth-like planets in the habitable zones (HZ) of their stars constitutes a fundamental science goal for a wide range of future scientific missions and concepts, and a central part of long-term scientific strategic plans of major astronomical research organizations ([Voyage 2050](); [Decadal Survey 2020]()). One of the most promising facilities in this context, which additionally is by far the most progressed at the time of writing, is the ESO ELT.

Several of the ELT instruments under development already include characterization of habitable worlds as part of their key science drivers. For example, (i) ANDES ([Palle et al. 2025]()) will feature high-resolution spectroscopy at visible and near-infrared wavelengths suitable for transmission spectroscopy of exoplanets orbiting in the HZs of M-dwarf; (ii) METIS ([Brandl et al. 2021]()) will feature imaging and medium- to high-resolution spectroscopy at mid-infrared wavelengths suitable for direct imaging and spectroscopy of HZ planets in the most nearby stellar systems; and (iii) PCS ([Kasper et al. 2021]()) will offer extreme adaptive optics at near-infrared wavelengths with sensitivity to HZ planets around primarily late-type stars.

These facilities will provide a great leap forward in our capacity to study planets in and near the HZ, particularly at high spectral resolution. However, in all of these cases, the instruments have a strong preference for super-Earths over Earths, as well as for late-type stellar hosts over Sun-like hosts. Such planets and systems feature vastly different conditions from those of Earth, even in cases where their equilibrium temperatures (which define the classical HZ) are very similar to that of Earth. For example, planets in the HZ around M-type stars are tidally locked, which will affect their atmospheric dynamics, and are subjected to much higher levels of high-energy radiation, stellar winds, and flares than planets in the HZ around Sun-like stars, all of which impact the planetary atmosphere and potentially the corresponding habitability ([Tilley et al. 2019](); [Zahnle et al. 2021]()).

In a very limited amount of individual cases, such as the alpha Cen system, the ELT instruments may probe also into the HZ of more Sun-like stars, which is the only range that we know for certain is genuinely habitable. However, the study of a *statistically relevant* sample of Earth-analog planets, which would be required to truly address the fundamental questions of habitability and life in the Universe that underpin the study of exoplanets, is beyond the presently planned capabilities even of the ELT. Fortunately, there is a way to reach such a sample and address these questions, and ELT is the best possible facility in the world to accomplish this – it just needs a bit of help from outer space.

**Starshades for direct exoplanet studies**

The great difference in brightness (high contrast) between a planet and its parent star renders the need to block out the stellar light to the greatest possible degree in order to study the planet. In the infrared, this can be accomplished though nulling interferometry, whilst at visible wavelengths, coronagraphs and starshades are the primary options. A coronagraph is placed inside of an instrument in order to block the starlight. Coronagraphic techniques are already being successfully used in direct studies of exoplanets, but reaching the contrast needed to detect Earth analogs places extremely high demands on the quality and stability of the telescope and its surrounding medium. A starshade, placed in space outside of the telescope, greatly alleviates these requirements by transmitting an already star-subtracted wavefront toward a telescope in its shadow. This enables a higher contrast over a wide passband than coronagraphy, with the drawback that switching between targets becomes significantly more complicated, given that the starshade each time needs to be navigated toward the new target at a considerable cost in fuel and time. Nonetheless, starshades can open scientific avenues that would simply be unreachable with coronagraphic techniques.

While starshade studies have classically focused on a starshade in space transmitting a shadow to a telescope that is also in space (e.g. [Copi & Starkman 2000](#); [Cash 2006](#); [Vanderbei et al. 2007](#)), several authors (e.g. [Janson 2007](#); [Soliman et al. 2025](#)) have noted that an even greater benefit could be acquired from configuring a starshade in space to transmit its shadow to a telescope on the surface of the Earth. A central problem in high-contrast imaging/spectroscopy from the ground is the fact that the atmosphere scrambles the wavefront from the bright star, effectively spreading flux from the core of its PSF into the wings where the planets reside. Although this effect can be mitigated with adaptive optics (AO), acquiring the wavefront correction required to reach Earth-Sun contrasts at separations of order 100 mas (1 au at 10 pc) is prohibitively challenging with present-day technology. If a starshade is placed in front of the star prior to the light entering the atmosphere, the disappearance of the star in the shadow of the starshade entirely alleviates the high-contrast nature of the problem. AO is still required in order to maximize the S/N of the planets and distinguishing them from zodiacal and exozodiacal light, but the corresponding requirements on the AO system are then entirely within the scope of already existing instrumentation.

The ability to keep a starshade shadow fixed (co-moving) on a specific point on Earth over 1-2 hour timescales, as well as being able to switch from one target to another in a different place on the celestial sphere, places very particular requirements on the starshade orbit. Still, several different families of orbits have been developed (e.g. [Janson 2007](#); [Janson et al. 2022](#); [Soliman et al. 2025](#)) that can meet these requirements under reasonable fuel budgets for the maneuvers required by an engine at the back of the starshade. The primary engineering challenge of the concept is to make the starshade (a mostly flat structure, of order 100 m across) as lightweight as possible, since that minimizes fuel consumption per boost required, and thus maximizes the number of targets that can be studied with the help of the starshade. Typical mission simulations imply that at least several tens of Sun-like target systems can be observed with a contrast sufficient to detect Earth analogs ($10^{-10}$) in the HZ, and hundreds of planets altogether. The telescope shadow where this level of contrast is provided is larger than the ELT, and the contrast is insensitive to meter-scale fluctuations in the position of the occulter and its shadow. With starshades being an important consideration for HWO, development of the technology is unavoidable, and can be leveraged for the purpose of an ELT starshade. More details about the various starshade mission concepts can be found in the provided references above.

The great advantage of relaying a starshade shadow to a telescope on the ground is that it provides access to much larger sized telescopes than what is possible or reasonable to launch into space. Furthermore, these telescopes can be equipped with much more sophisticated and easily serviceable instrumentation, such as high-resolution spectrographs. Even with already fully functional facilities, such as e.g. VLT/MUSE ([Bacon et al. 2010](#)), it would be possible to acquire high-fidelity results on Earth analogs ([Janson et al. 2022](#)) with the help of a starshade. If the ELT is combined with a starshade, a thoroughly unique realm of achievable sensitivity to HZ planets opens up: Launching a 39m-class telescope into space would be an impossible task; yet, with the starshade concept, the ELT suddenly acquires the ability to study Earth analogs with a comparable accuracy as if it were in space. Telluric absorption does still provide an effect in some wavelength ranges that would not occur in space, but is non-critical for the most important biomarker molecules even at low spectral resolution ([Janson et al. 2022](#)). With high spectral resolution, the telluric and target absorption lines can be separated in velocity space, effectively alleviating the issue entirely.

**Optimizing the synergy between ELT instrumentation and a starshade**

Importantly, starshades only produce contrasts useful for Earth analog purposes in the visible wavelength regime. The wavelength regime covered by a starshade depends on its exact design, but as a general rule of thumb, 1 μm is the upper cut-off wavelength for a $10^{-10}$ level contrast performance, with very quick deterioration for longer wavelengths. Meanwhile, observations of

Earth analogs require AO. The combination of visible light and AO has classically been unusual, due to the fact that it is more difficult to provide high-quality wavefront correction at visible than at IR wavelengths. However, with technological improvements, visible-light AO has become increasingly used, and forms an important component of present-day instrumentation such as in MUSE and SPHERE-ZIMPOL. An AO system with as powerful performance as possible in the visible wavelength range is thus desirable. When the target star is hiding behind the starshade during observations, it cannot act as guide star for the AO, but the starshade can be equipped with a laser that forms a guide star in a specific constrained wavelength range. The operating range of the laser and the operating range of the AO system obviously need to be mutually compatible.

Unless the planets to be studied are known in advance with precisely determined ephemerides, the ELT instrument used should preferable be an integral field unit (IFU), such that a spectrum can be acquired for any planet visible in the field of view during starshade observations, regardless of where it resides within the field. This requires an IFU with a sufficient field of view to cover any possible angular separations from the star that the planet might reasonably have. For an Earth analog in the α Cen system, the maximum separation is approximately 750 mas, though this is rather extreme – for other systems, the maximum separation would be considerably smaller. On the other hand, when given the chance to observe planetary systems with the star blocked out by the starshade, it is of scientific interest to fully characterize the system by also studying outer planets – for example analogs of Mars, Jupiter, Saturn, etc. For these reasons, a relatively wide field of view is always desirable, although separations beyond about 750 mas are never critical for the key science case of Earth analogs. Meanwhile, the spatial sampling should not be too coarse. If the size of a pixel is much wider than the FWHM of the planet PSF, there will be an unnecessarily large contribution from noise sources encompassed in the pixel. A Nyqvist sampling of the PSF size would, as always, be the ideal case, but a pixel scale as coarse as 50 mas could be workable for nearby targets, depending on exozodi levels and leakage flux levels outside the inner working angle (IWA).

While useful exoplanet characterization studies could be executed at any spectral resolution (or even with pure imaging), high spectral resolution offers some unmatchable advantages. A high resolution (R = 100,000 is ideal) enables a clean separation between telluric features and exoplanetary features, and provides quantities such as the radial and rotational velocities of the planet, which would be otherwise unattainable. Cross-correlation / molecular mapping techniques can be used to extract high-quality spectral information even in circumstances where the S/N per wavelength bin is low, as frequently demonstrated in contemporary studies (Snellen et al. 2014).

Among currently planned instrumentation for ELT, HARMONI (under current scoping) includes many of the required features of ELT+starshade synergy such as an AO-assisted IFU covering 470-1000 nm (and longer), with spatial scales of 4-20 mas at FOVs of about 0.6-3". A primary potential improvement for enhancing the synergy is an increased spectral resolution (currently 3500 in VIS). ANDES offers the highly tantalizing opportunity of R = 100,000 in VIS. Adapting it for starshade synergy would involve extending AO capacity (NIR under current scoping) to include VIS, and preferably increasing the number of spaxels (current scoping 7x7) to about 32x32, which could allow for (e.g.) 1.6" FOV with 50 mas sampling for the nearest target systems, and 0.4" FOV with 12.5 mas sampling for the most distant ones. For MICADO, synergy with a starshade would be improved by extending the VIS AO capability shortward of the current lower cut-off (800 nm), as well as pushing the spectral resolution as high as possible (currently scoped at 10,000-20,000). Any other extensions of ELT instruments into the VIS range, or specialised ELT instruments for starshade synergy, should strive for similar requirements as those outlined above.